\documentclass[aps, prl, letterpaper, amsmath, amssymb, preprintnumbers, showpacs, superscriptaddress, twocolumn, floatfix,, nofootinbib]{revtex4-1}
\usepackage{xspace}
\usepackage{graphicx}
\usepackage{subfigure}
\usepackage{float}
\usepackage{color}
\usepackage{comment}
\usepackage[dvipsnames]{xcolor} \pdfoutput=1
\allowdisplaybreaks 
\usepackage[hypertexnames=false]{hyperref}
\hypersetup{
    colorlinks=true,       
    linkcolor=blue,          
    citecolor=blue,        
    filecolor=blue,      
    urlcolor=blue           
}
\usepackage{slashed}
\usepackage{mathtools}

\newcommand{\hethree}{{}^3\text{He}}

\newcommand{\mbraket}[3]{\left< #1 \vphantom{#2#3} \right|
 #2 \left| #3 \vphantom{#1#2} \right>} 

\begin{document}

\begin{figure}
  \vskip -1.5cm
  \leftline{\includegraphics[width=0.15\textwidth]{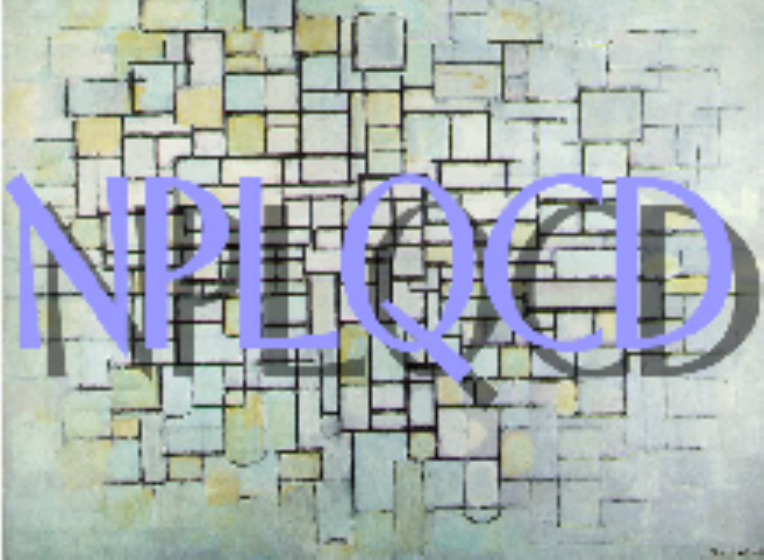}}
\end{figure}

\title{Lattice QCD constraints on the parton distribution functions of $\hethree$}

\author{William Detmold} \affiliation{
	Center for Theoretical Physics, 
	Massachusetts Institute of Technology, 
	Cambridge, MA 02139, USA}

\author{Marc Illa}
\affiliation{Departament de F\'{\i}sica Qu\`{a}ntica i Astrof\'{\i}sica and Institut de Ci\`{e}ncies del Cosmos,
	Universitat de Barcelona, Mart\'{\i} Franqu\`es 1, E08028-Spain}

\author{David J. Murphy}  \affiliation{
	Center for Theoretical Physics, 
	Massachusetts Institute of Technology, 	Cambridge, MA 02139, USA}

\author{Patrick Oare } \affiliation{
	Center for Theoretical Physics, 
	Massachusetts Institute of Technology, 
	Cambridge, MA 02139, USA}

\author{Kostas~Orginos}
\affiliation{Department of Physics, College of William and Mary, Williamsburg, VA 23187-8795, USA}
\affiliation{Jefferson Laboratory, 12000 Jefferson Avenue, Newport News, VA 23606, USA}

\author{Phiala E. Shanahan } \affiliation{
	Center for Theoretical Physics, 
	Massachusetts Institute of Technology, 
	Cambridge, MA 02139, USA}

\author{Michael L. Wagman}
\address{Fermi National Accelerator Laboratory, Batavia, IL 60510, USA}

\author{Frank Winter}
\affiliation{Jefferson Laboratory, 12000 Jefferson Avenue, Newport News, VA 23606, USA}

\collaboration{NPLQCD collaboration}

\begin{abstract}
The fraction of the longitudinal momentum of $\hethree$ that is carried by the isovector combination of $u$ and $d$ quarks is determined using lattice QCD for the first time.
The ratio of this combination to that in the constituent nucleons is found to be consistent with unity at the few-percent level from calculations with quark masses corresponding to $m_\pi\sim 800$~MeV, extrapolated to the physical quark masses.
This constraint is consistent with, and significantly more precise than, determinations from global nuclear parton distribution function fits. Including  the lattice QCD  determination of the momentum fraction in the nNNPDF global fitting framework results in the uncertainty on the isovector momentum fraction ratio being reduced by a factor of 2.5, and thereby enables a more precise extraction of the $u$ and $d$ parton distributions in $\hethree$.
\end{abstract}
\preprint{MIT-CTP/5234, ICCUB-20-019, FERMILAB-PUB-20-466-T}
\maketitle


A central pillar of our understanding of the internal structure of strongly interacting hadronic and nuclear systems is knowledge of their partonic structure as accessed in deep-inelastic scattering (DIS) experiments and other hard processes. Since the 1960s, such experiments have revealed the longitudinal momentum distributions of quarks and gluons in a fast moving proton, known  collectively as parton distribution functions (PDFs). The simplest PDFs, $q(x,\mu)$ (and $g(x,\mu)$),  describe the probability of a quark of flavor $q$ (or gluon $g$) carrying a fraction $x$ of the longitudinal momentum of the  struck proton at a renormalization scale $\mu$. In 1983, the European Muon Collaboration (EMC) \cite{Aubert:1983xm}  observed that the partonic structure of nuclei differs substantially from that of the constituent protons and neutrons, a landmark in the development of nuclear physics  \cite{Arneodo:1992wf,Geesaman:1995yd,Piller:1999wx,Norton:2003cb,Hen:2013oha}. Since the DIS processes observed in the EMC experiments were at very high energy, and the binding energy of a nucleus is small in comparison to its mass,  the appearance and size of 
the EMC effect was surprising at the time. Interest in the EMC effect has been rekindled by recent data from SLAC and JLab \cite{Fomin:2011ng,Hen:2012fm,Frankfurt:1993sp,Egiyan:2003vg,Egiyan:2005hs} on EMC ratios for light nuclei. Not only have these data provided precise determinations of the EMC effect for nuclei with small atomic number $A$, but they have revealed a correlation between the strength of the EMC effect and so called ``short range correlations'' \cite{Weinstein:2010rt,hen2012}. 

In addition to experimental investigations, theoretical calculations of the partonic structure of hadrons and nuclei from the Standard Model can have important impact on our understanding of the structure of matter. 
For example, Standard Model calculations of nuclear partonic structure would reveal the QCD origin of the EMC effect as well as aid in the flavor-separation of proton PDFs. Parton distributions are inherently rooted in the strong interaction dynamics of QCD and cannot be determined using perturbative methods. Since the seminal works of Refs. \cite{Martinelli:1987bh,Martinelli:1988rr}, lattice Quantum Chromodynamics (LQCD) calculations have addressed the simplest aspects of the parton distributions of the proton, notably determining the first few Mellin moments of the unpolarized, polarized, and transversity quark distributions \cite{Lin:2017snn}, as well as their gluonic analogues \cite{Horsley:2012pz,Alexandrou:2016ekb,Yang:2018bft,Shanahan:2018pib,Shanahan:2018nnv}. Recently, efforts have been made to extend these studies to the full $x$-dependence of the proton PDFs \cite{Ji:2013dva,Lin:2017snn,Cichy:2018mum,Ji:2020ect}. More complicated extensions of partonic structure, such as generalized parton distribution functions and transverse-momentum dependent parton distribution functions of the proton, have also been studied using LQCD~\cite{Hagler:2009ni,Yoon:2017qzo,Shanahan:2020zxr,Alexandrou:2020zbe,Lin:2020rxa,Zhang:2020dbb}.  

In this letter, the partonic structure of light nuclei  is studied in LQCD for  the  first  time  through  an  investigation of the  isovector quark  momentum fractions (the first moments of the corresponding isovector PDFs) of the proton, diproton, and $\hethree$. At the heavier-than-physical quark masses used in this LQCD study, percent-level nuclear effects are resolved in the momentum fraction of $\hethree$. After an extrapolation to the physical quark masses, these calculations provide a constraint on the isovector momentum fraction that is used as an additional input into the nNNPDF2.0 \cite{AbdulKhalek:2020yuc} global nuclear PDF analysis framework. Since the isovector combination of nuclear PDFs is poorly determined from experiment, this LQCD constraint significantly reduces the uncertainties on the $\hethree$ PDFs and thereby improves knowledge of nuclear structure.

\textbf{\textit{LQCD methodology --- }}
The existence of strong interactions between quarks and gluons necessitates the use of LQCD for calculations of the partonic structure of nuclei. 
The calculations presented here are performed using  a single ensemble of gauge-field configurations generated with a L{\"u}scher-Weisz gauge action~\cite{Luscher:1984xn} with $N_f = 3$ degenerate light-quark flavors with the clover-improved Wilson fermion action~\cite{Sheikholeslami:1985ij}, and quark masses tuned to produce a pion mass of $m_\pi = 806$ MeV.
The lattice geometry is $L^3\times T =32^3\times 48$, and the lattice spacing is determined to be $a \sim 0.145$ fm from $\Upsilon$ spectroscopy~\cite{Beane:2012vq}.
This ensemble, and two others with different spacetime volumes have previously been used to study the spectrum \cite{Beane:2012vq,Wagman:2017tmp} and properties \cite{Beane:2017edf,Beane:2013br,Beane:2014ora,Beane:2015yha,Chang:2015qxa,Detmold:2015daa,Savage:2016kon,Tiburzi:2017iux,Shanahan:2017bgi,Winter:2017bfs,Chang:2017eiq} of light nuclei up to atomic number $A=4$. The multi-volume spectroscopy studies show that the $pp$ and $\hethree$ states that are investigated here are bound systems with infinite volume energies below threshold. Consequently, matrix elements in these states are expected to receive only exponentially small finite volume effects, ${\cal O}(e^{-\kappa L}, e^{-m_\pi L})$, that will be neglected in this work~\cite{Luscher:1985dn,Luscher:1986pf,Luscher:1990ux,Beane:2003da,Davoudi:2011md,Konig:2017krd,Briceno:2019nns}. 

The Mellin moments of the unpolarized isovector quark PDFs, $q_3^{(h)}(x,\mu)=u^{(h)}(x,\mu)-d^{(h)}(x,\mu)$, in a hadronic or nuclear state $h$, defined as $\langle x^n \rangle_{u-d}^{(h)}(\mu) \equiv \int_{-1}^{1} dx \; x^n q_3^{(h)}(x,\mu)$, are determined from matrix elements of  twist-two operators as
\begin{align}
\langle h |{\cal O}_{\mu_0\ldots\mu_n} | h\rangle & \equiv
\langle h | \overline q  \tau_3 \gamma_{\{\mu_0} (i\overleftrightarrow{D}_{\mu_1})\ldots(i \overleftrightarrow{D}_{\mu_n\}})q| h \rangle \nonumber \\
&=\langle x^n \rangle_{u-d}^{(h)}(\mu) p_{\{\mu_0}\ldots p_{\mu_n\}}, 
\label{eq:opdef}
\end{align}
where $p$ is the momentum of the state $h$, $\tau_3$ is a Pauli matrix in flavor space, $\overleftrightarrow{D}_{\mu} =(\overrightarrow{D}_{\mu} - \overleftarrow{D}_{\mu})/2$ where $D_\mu$ is the gauge covariant derivative, and $\{\ldots\}$ indicates symmetrization and trace-subtraction of the enclosed indices. 
The above operators are constructed to transform irreducibly under the Lorentz group, but the hypercubic spacetime lattice used in the LQCD calculations reduces these symmetries, in general inducing mixing between operators of different Lorentz spin.
In particular, the two-index operators that determine the isovector quark momentum fraction, $\langle x\rangle_{u-d}^{(h)}$, subduce to operators in two different irreducible representations of the hypercubic group. In this work, matrix elements of an Euclidean operator in the  $\tau_1^{(3)}$ representation~\cite{Gockeler:1996mu} are computed, namely 
\begin{equation}
{\cal T} = {{\frac{1}{\sqrt{2}}}} \left( {\cal T}_{33} - {\cal T}_{44} \right),\:\ {\rm with} \:\ {\cal T}_{\mu\nu} = \overline{q} \tau_3\gamma_{\{\mu}\overleftrightarrow{D}_{\nu\}}  q\,,
\label{eq:Tqlattdef}
\end{equation}
where $\gamma_\nu$ is also Euclidean.
With a lattice regulator, this operator is discretized as a covariant finite difference whose form is given in the Supplementary Material.
For both spin-zero and spin-half systems, spin-averaged in the latter case,  matrix elements in states with zero three-momentum determine the momentum fraction as
$\mbraket{h}{{\cal T}}{h} = 
   \langle x\rangle_{u-d}^{(h)} M_h/\sqrt{2}$.

The renormalized operator in the modified minimal subtraction scheme ($\overline{\rm MS}$) is related to the bare lattice operator in Eq.~\eqref{eq:Tqlattdef} as
\begin{equation}
    {\cal T}^{(\overline{\rm MS})} (\mu)=
	\mathcal{R}^{\overline{\rm MS}/\rm{RI'MOM}}(\mu,\mu_0)	\mathcal{Z}^{\rm{RI'MOM}}(\mu_0,a)  {\cal T}(a),
    \label{eq:renorm}
\end{equation}
where the renormalization coefficient ${\cal Z}^{\rm RI'MOM}(\mu_0,a)$ is defined  non-perturbatively in a regularization-independent momentum-subtraction scheme \cite{Martinelli:1994ty} at a scale $\mu_0$ and then matched to  $\overline{\rm MS}$ through the three-loop perturbative coefficient ${\cal R}^{\overline{\rm MS}/{\rm RI'MOM}}(\mu,\mu_0)$ \cite{Gracey:2003mr,Gracey:2003yr}, as detailed in the Supplementary Material. For $\mu=2$ GeV,  $ {\cal R}^{\overline{\rm MS}/{\rm RI'MOM}}(\mu,\mu_0) {\cal Z}^{\rm RI'MOM}(\mu_0,a)=0.89(4)$.

The  techniques needed to compute matrix elements of this operator are simple generalizations of those used for calculations of isovector matrix elements of quark currents using the compound-propagator background-field method introduced in Ref.~\cite{Savage:2016kon} and further detailed in Refs.~\cite{Bouchard:2016heu,Shanahan:2017bgi,Tiburzi:2017iux} and the Supplementary Material. 
Quark propagators and ${\cal T}$-compound propagators are computed from an average of $N_{\rm src} = 24$ source points randomly distributed on  $N_{\rm cfg} = 2290$ gauge-field configurations for $N_{B}=5$ different background field strengths.
These compound propagators are then used to construct baryon two-point correlation functions,
\begin{equation}
      G_h(t;\lambda) = \sum_{\mathbf{x}} \left\langle 0\left|\chi_h(\mathbf{x},t) \chi^\dagger_h(0) \right|0\right\rangle_\lambda\,,
\label{eq:G2ptdefBF}
\end{equation}
where  $\lambda$ is the ${\cal T}$-background field strength, $\chi_h$ is an interpolating field for states with the quantum numbers of the hadron or nucleus $h$, and spinor indices on the interpolating operators are suppressed.
Correlation functions are constructed from  Gaussian-smeared source interpolating operators  \cite{Albanese:1987ds}, while the sink interpolating operators are either smeared or point-like and the multi-baryon contractions are performed using the techniques of Refs.~\cite{Detmold:2012eu}.  
This quantity contains responses to the field up to ${\cal O}(\lambda^{N_Q})$, with $N_Q$ being the number of valence quarks in the state. The linear response of this background-field two-point function, $\left. G_h(t;\lambda)\right|_{{\cal O}(\lambda)}$, is determined by the matrix element of ${\cal T}$. This term can be extracted exactly from the computed set of fixed-order background field correlation functions with $N_{Q}$ field strengths \cite{Savage:2016kon,Tiburzi:2017iux}.

Combining the linear response of the two-point correlation function with the zero-field correlation function, it is straightforward to show that the ratio
\begin{equation}
      \mathcal{R}_h(t) = \frac{\left. G_h(t;\lambda)\right|_{{\cal O}(\lambda)}}{G_h(t;0)} - \frac{\left. G_h(t-a;\lambda)\right|_{{\cal O}(\lambda)}}{G_h(t-a;0)}
\label{eq:R3ptdef}
\end{equation}
is related to matrix elements of ${\cal T}$ through the spectral representation of each term in Eq.~\eqref{eq:R3ptdef}, in particular asymptoting as
\begin{equation}
      \mathcal{R}_h(t) \overset{t\to\infty}{\longrightarrow} \mbraket{h}{{\cal T}}{h}, 
\label{eq:R3ptspec}
\end{equation}
with exponentially vanishing contamination at early times that involves excited-state overlap factors and transition matrix elements.
\begin{figure}[t!] 
  \centering
  \includegraphics[width=0.475\textwidth]{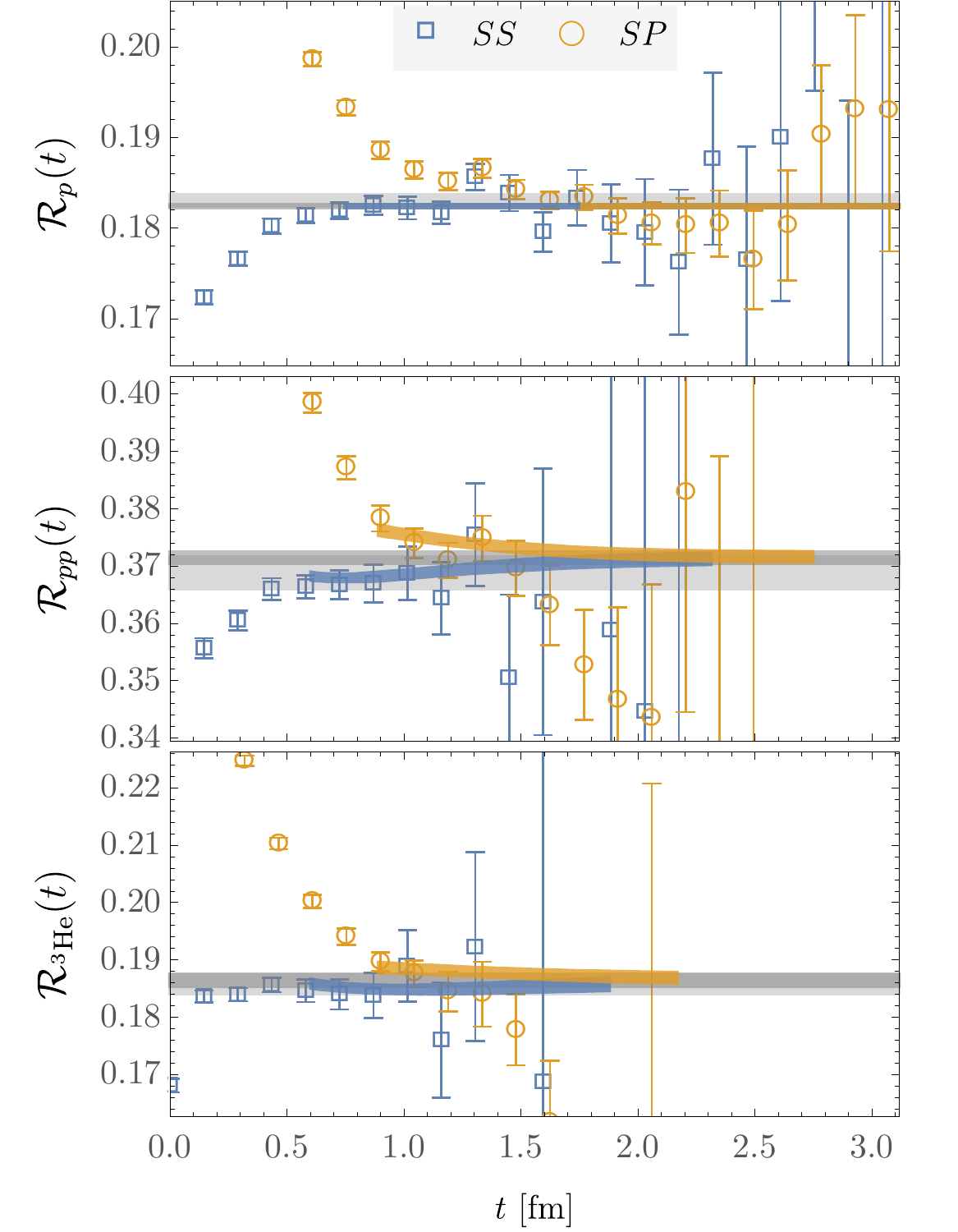}
  \caption{
     The effective matrix element, Eq.~\eqref{eq:R3ptdef}, associated with the isovector quark momentum fractions of the proton, $pp$ and $\hethree$. Blue (orange) points, labelled SS (SP), show results for interpolating operators with smeared sources and smeared (point-like) sinks. 
     For each effective matrix element, points are shown for $t \leq t_{\rm max}$, where $t_{\rm max}$ is the minimum $t$ where the signal-to-noise ratio of $G_h(t+a;\lambda)|_{\mathcal{O}(\lambda)}$ is less than 0.5.
     Colored bands show the highest weight fit to the combined dataset and the shaded gray bands show the weighted average of all accepted fits and the total statistical plus fitting systematic uncertainties.
    \label{fig:Rbar_bestfit} }
\end{figure}

Ground-state matrix elements are extracted from ${\cal R}_h(t)$, and systematic fitting uncertainties are estimated, using a procedure for sampling from all possible fit ranges and models analogous to the procedure described for two-point correlation functions in Ref.~\cite{Beane:2020ycc}.
In summary, in analyzing ${\cal R}_h(t)$ to extract the momentum fractions, the full $t$ dependence that results from the spectral decomposition of each term in Eq.~\eqref{eq:R3ptdef} is fit, and combined fits to two- and three-point correlation functions are used to constrain the relevant energies, overlap factors, and matrix elements. All possible choices of fit ranges and up to 4 states contributing to the spectral decompositions are considered using a model selection process described in the Supplementary Material. A weighted average over fits from all acceptable fit ranges is used to define ground-state energy results, including systematic uncertainties from fit range and model variation.
Results are shown in Fig. \ref{fig:Rbar_bestfit} for the proton, diproton and $\hethree$.

\begin{figure}[!t]
  \centering
    \vspace{-2mm}
  \includegraphics[width=0.45\textwidth]{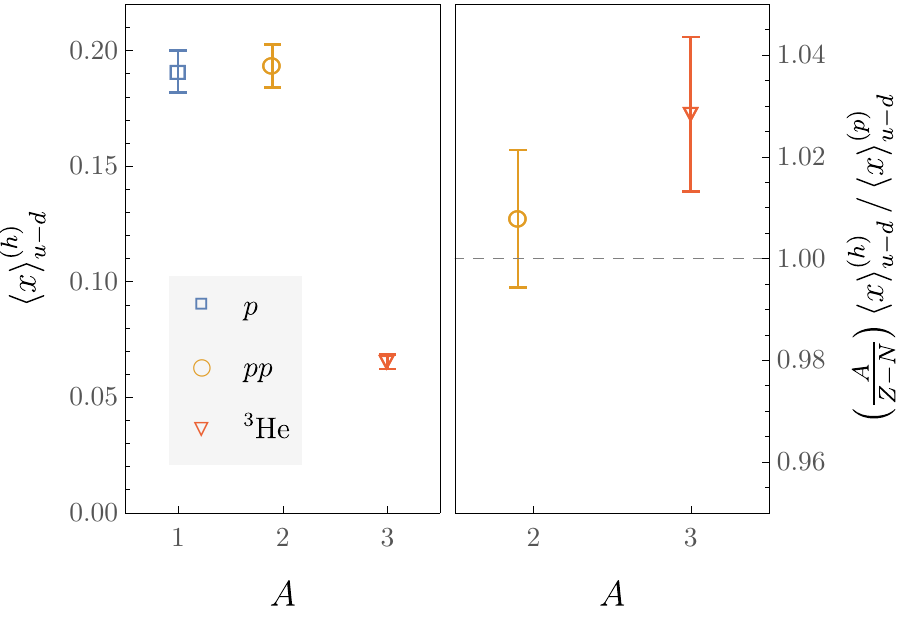}
  \vspace{-5mm}
  \caption{
     Left: Renormalized isovector momentum fractions for  $h \in \{p, pp,\hethree\}$ at a scale of $\mu=2$~GeV.
     Right: Ratios of the isovector nuclear momentum fractions to that of the constituent nucleons. 
    \label{fig:momfrac_summary}}
\end{figure}

\textbf{\textit{Results and Discussion --- }}\label{sec:operators}
The extracted values of the isovector quark momentum fractions for $p,\,pp,\,\hethree$ at quark masses corresponding to $m_\pi=m_K=806$ MeV are shown in Tab.~\ref{tab:bare} and displayed graphically in Fig.~\ref{fig:momfrac_summary}. The uncertainties are separated into those from the LQCD calculation of the bare matrix elements, and the (larger) uncertainty from the renormalization and matching to the $\overline{\rm MS}$ scheme. The proton isovector momentum fraction is consistent with other LQCD extractions at similar values of the quark masses \cite{Dolgov:2000ca} given the different renormalization procedures and lattice spacings. The $pp$ and $\hethree$ momentum fractions are determined with ${\cal O}(5\%)$ uncertainties and are found to be approximately consistent with those of the constituent nucleons. The ratios of the nuclear momentum fractions to that of the proton are independent of operator renormalization to ${\cal O}(\alpha_s)$, and are determined at few-percent precision even for $\hethree$. 
\begin{table}[!t]
\begin{ruledtabular}
\begin{tabular}{cccc}
            & $p$                   & $pp$          & $\hethree$     \\\hline
            $ \langle x \rangle_{u-d}^{(h)}$   & $0.191(1)(9)$         & $0.194(2)(9)$  & $0.066(1)(3)$   \\
   $ \left( \frac{A}{Z-N} \right) \langle x \rangle_{u-d}^{(h)}/\langle x \rangle_{u-d}^{(p)}$   & ---       & $1.007(14)$  & $1.028(15)$   \\
  \end{tabular}
  \caption{The isovector quark momentum fractions in  $p$, $pp$ and $\hethree$, calculated at $m_\pi=806$~MeV in $\overline{\rm MS}$-scheme at $\mu=2$ GeV. The first uncertainty combines LQCD statistical and systematic uncertainties and the second uncertainty is from operator renormalization. The correlated ratios of the isovector momentum fraction in nuclei to those in the constituent nucleons, in which the renormalization constants and their uncertainties cancel, are also given. 
  \label{tab:bare}}
   \end{ruledtabular}
\end{table}

In Refs.~\cite{Chen:2004zx,Chen:2016bde,Lynn:2019vwp}, nuclear effective field theory (EFT) was used to study nuclear effects in PDF moments. In particular, it was shown that the leading source of such effects is the two-nucleon correlations that couple to the twist-two operators defining the PDF moments. In terms of the parameters defined in that work, nuclear effects in the isovector momentum fraction are encapsulated in the low energy constant (LEC) $\alpha_{3,2}$ and nuclear factor ${\cal G}_3(\hethree)$; their product is bounded as $\alpha_{3,2}{\cal G}_3(\hethree)=0.0018(14)$ at $\mu=2$ GeV from the numerical calculations presented here (see the Supplementary Material for details). While the quark momentum fractions themselves have nonanalytic dependence on the quark masses \cite{Detmold:2001jb,Arndt:2001ye,Chen:2001eg}, this two-body LEC is expected to be relatively insensitive to variation of the quark masses, as seen for the the analogous two-body contribution in  the $np\to d\gamma$ \cite{Beane:2015yha} and $pp\to d e^+\nu_e$ \cite{Savage:2016kon,TOAPPEAR} processes.  
This relative mass-independence assumption allows an extrapolation to the physical quark masses: a naive estimate is given by taking the central value determined at $m_\pi=806$ MeV and inflating the uncertainty by 50\% to account for possible quark-mass dependence as well as the effects of the nonzero lattice spacing and finite volume (this uncertainty is estimated based on the mass dependence seen for the analogous two-body LECs in Refs.~\cite{Beane:2015yha,Savage:2016kon,TOAPPEAR}). This  extrapolated value can be combined with the physical value of the nucleon momentum fraction, $\langle x\rangle_{u-d}^{(p)}=0.160(7)$  at $\mu=2$ GeV from the nNNPDF2.0 analysis~\cite{AbdulKhalek:2020yuc}, to determine the  isovector momentum fraction ratio $3 \langle x \rangle_{u-d}^{(^3\text{He})}/\langle x \rangle_{u-d}^{(p)}|_{\text{LQCD}}=1.035(26)$ at the physical quark masses (see the Supplementary Material for more details).

\begin{figure}[!t] 
  \centering
  \includegraphics[width=0.44\textwidth]{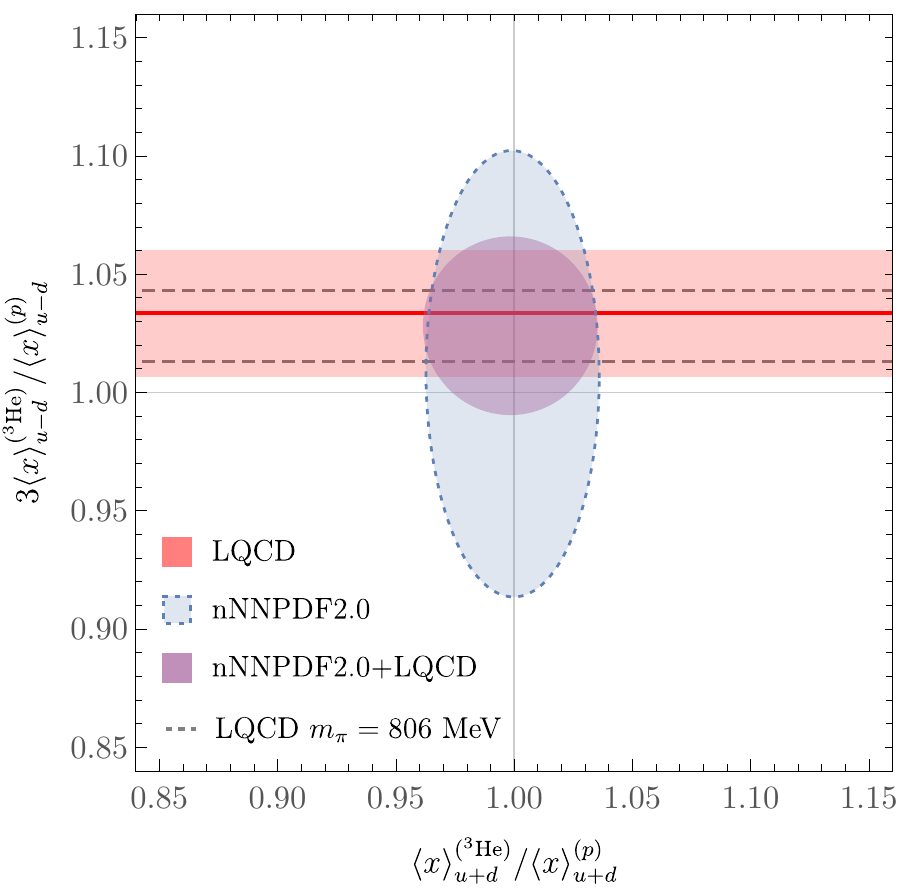}
  \caption{
     The ratio of the isovector momentum fractions of ${}^3{\rm He}$ and $p$ determined in this work compared to constraints on the isovector and isoscalar momentum fraction ratios from the nNNPDF2.0 \cite{AbdulKhalek:2020yuc}  global analysis before and after the LQCD constraint is imposed. Both axes are normalized to unity in the absence of nuclear effects. The  LQCD constraint on the isovector ratio at $m_\pi=806$ MeV is also displayed. In all cases, 68\% confidence intervals are shown. 
    \label{fig:momfrac_comparison}
    }
\end{figure}

It is interesting to compare the LQCD results for the momentum fractions and their ratios to  phenomenology. In particular, the isovector momentum fractions determined here provide valuable information that is complementary to experimental constraints on the nuclear modification of PDFs; almost all information on the nuclear modification of partonic structure has been obtained for the ratio of isoscalar-corrected $F_2$ structure functions of nuclei to that of the deuteron \cite{Geesaman:1995yd,Norton:2003cb,Hen:2013oha}. 
Additional constraints are especially valuable in the context of the intriguing question as to whether there is flavor-dependence to the EMC effect. Such flavor dependence has been conjectured in models of QCD \cite{Uchiyama:1988xs,Afnan:2000uh,Saito:2000fx,Bentz:2009yy,Cloet:2012td,Tropiano:2018quk} and in EFT \cite{Chen:2004zx,Chen:2016bde,Lynn:2019vwp} and is included in recent data-driven analyses of experimental results \cite{Schmookler:2019nvf,Segarra:2019gbp} and provides a potential explanation of the NuTeV anomaly in $\sin^2\theta_W$ \cite{Cloet:2009qs}. 

\begin{figure}[!t] 
  \centering
  \includegraphics[width=0.45\textwidth]{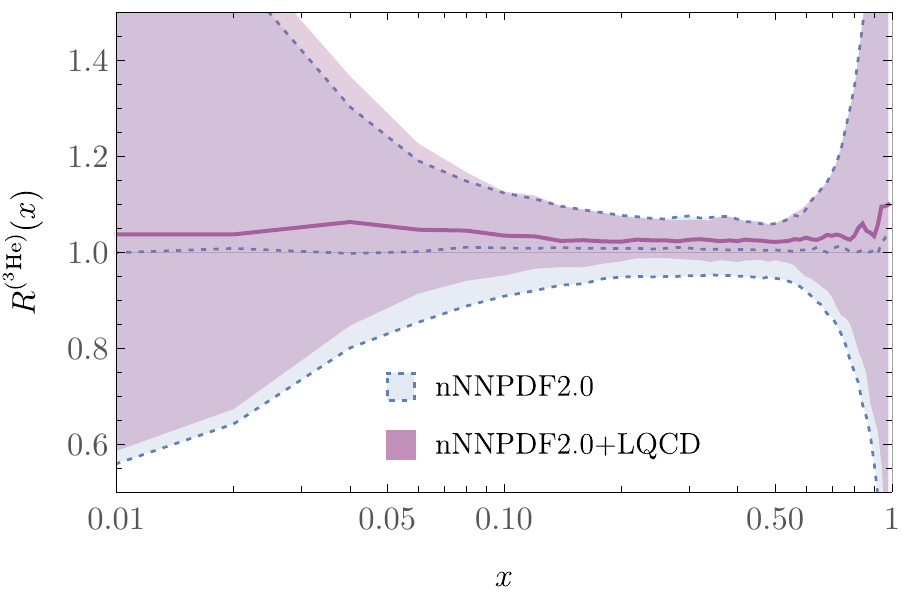}
  \caption{
      The ratio $R^{(\hethree)}(x)=3q_3^{(\hethree)}(x)/q_3^{(p)}(x)$ of the nNNPDF2.0 isovector  PDF in $\hethree$ to that in the proton  \cite{AbdulKhalek:2020yuc}, as well as the same distribution with the 
      LQCD moment constraint imposed into the global analysis as described in the text. 68\% confidence intervals are shown.
    \label{fig:pdf_comparison}}
\end{figure}

Fig.~\ref{fig:momfrac_comparison} shows the constraint on the isovector momentum fraction ratio for $\hethree$ obtained from the results presented here, compared with the constraints  on the isovector and isoscalar momentum fraction ratios from the recent nNNPDF2.0 \cite{AbdulKhalek:2020yuc} global nuclear PDF fits. The nNNPDF2.0 ellipse is generated by combining the Monte Carlo replica sets for the bound proton PDFs in $^4$He appropriately to form the PDFs of $^3$He (under the assumption that the nuclear effects vary slowly with $A$). In this way, correlations between the $\hethree$ and proton PDFs are accounted for. For the isovector combination, the 68\% confidence interval is $3 \langle x \rangle_{u-d}^{(^3\text{He})}/\langle x \rangle_{u-d}^{(p)}|_{\text{nNNPDF2.0}}=1.007(63)$.
In the nNNPDF approach, it is also straightforward to impose the LQCD constraint on the nuclear PDFs by reweighting the Monte Carlo replicas as discussed in Ref.~\cite{Ball:2011gg}; the combined confidence region is shown in Fig.~\ref{fig:momfrac_comparison}.  The 68\% confidence interval reduces to $3 \langle x \rangle_{u-d}^{(^3\text{He})}/\langle x \rangle_{u-d}^{(p)}|_{\text{nNNPDF2.0+LQCD}}=1.028(25)$.
Fig. \ref{fig:pdf_comparison} compares the ratio of the isovector PDF for $\hethree$  to that of the constituent nucleons, with and without the imposition of the LQCD constraint. 
As can be seen from the reduced uncertainties in Figs.~\ref{fig:momfrac_comparison} and \ref{fig:pdf_comparison}, LQCD calculations such as those presented here, as well as new experimental constraints~\cite{Mousseau:2016snl,MARATHON}, can significantly improve our knowledge of the flavor dependence of nuclear PDFs.

\textbf{\textit{Summary --- }} 
In this work, the isovector momentum fractions of the proton, diproton and $\hethree$ systems have been determined using LQCD, complementing a previous study of the gluon momentum fraction on the same ensemble \cite{Winter:2017bfs}. 
These calculations were performed at a single set of unphysical SU(3)-symmetric values for the quark masses corresponding to $m_\pi=806$ MeV, and in a single lattice volume and at a single lattice spacing.
Bearing these caveats in mind, the isovector nuclear momentum fractions were calculated precisely and found to be similar to that of the proton. In particular, the ratios    $ \langle x \rangle_{u-d}^{(pp)}/\langle x \rangle_{u-d}^{(p)}=1.010(14)$ and    $3 \langle x \rangle_{u-d}^{(\hethree)}/\langle x \rangle_{u-d}^{(p)}=1.029(15)$ were determined and nuclear EFT arguments were used to connect the $\hethree$ result to global analyses of nuclear PDFs, providing important constraints on the flavor decomposition of nuclear PDFs that are complementary to those obtained from experiment.

While in its early stages, this work emphasizes the utility of LQCD in constraining less well-measured aspects of partonic structure in an analogous  way to how LQCD inputs have been used to constrain the proton transversity PDFs \cite{Lin:2017stx}.
Future calculations  at the physical quark masses will consider higher moments of nuclear PDFs (or even directly study their $x$ dependence) for a wider range of nuclei and provide a complete flavor decomposition. Calculations will also quantitatively address the full set of systematic uncertainties.

\textbf{\textit{Acknowledgements --- }}
We thank Juan Rojo for comments and substantial help with the comparison to the nNNPDF2.0 global fits, and Silas Beane, Jiunn-Wei Chen, Zohreh Davoudi, Assumpta Parre\~no, Martin Savage and Brian Tiburzi for  insightful discussions.
This research used resources of the Oak Ridge Leadership Computing Facility at the Oak Ridge National Laboratory, which is supported by the Office of Science of the U.S. Department of Energy under Contract number DE-AC05-00OR22725, as well as facilities of the USQCD Collaboration, which are funded by the Office of Science of the U.S. Department of Energy. This research used resources of the National Energy Research Scientific Computing Center (NERSC), a U.S. Department of Energy Office of Science User Facility operated under Contract No. DE-AC02-05CH11231. The Chroma~\cite{Edwards:2004sx}, Qlua~\cite{qlua}, QUDA~\cite{Clark:2009wm,Babich:2010mu}, QDP-JIT \cite{Winter:2014npa} and QPhiX~\cite{10.1007/978-3-319-46079-6_30} software libraries were used in data production and analysis. 
WD, DJM and PES acknowledge support from the U.S.~DOE grant DE-SC0011090.
WD is also supported within the framework of the TMD Topical Collaboration of the U.S.~DOE Office of Nuclear Physics, and by the SciDAC4 award DE-SC0018121.
PES is additionally supported by the National Science Foundation under CAREER Award 1841699 and under EAGER grant 2035015, by the U.S. DOE Early Career Award DE-SC0021006, by a NEC research award,  by the Carl G and Shirley Sontheimer Research Fund.
MI is supported by the Universitat de Barcelona through the scholarship APIF, by the Spanish Ministerio de Econom\'{\i}a y Competitividad (MINECO) under the project No. MDM-2014-0369 of ICCUB (Unidad de Excelencia “Mar\'{\i}a de Maeztu”) and with additional European FEDER funds under the contract FIS2017-87534-P. 
KO was supported in part by U.S.~DOE grant \mbox{
DE-FG02-04ER41302} and in part by the Jefferson
Science Associates, LLC under U.S.~DOE Contract DE-AC05-06OR23177.  
This manuscript has been authored by Fermi Research Alliance, LLC under Contract No. DE-AC02-07CH11359 with the U.S. Department of Energy, Office of Science, Office of High Energy Physics. The authors thank Robert Edwards, B\'{a}lint Jo\'{o}, and members of the NPLQCD collaboration for generating and allowing access to the ensembles used in this study, as well as for helpful discussions.

\bibliography{momfrac_bib}

\newpage
\clearpage
\newpage
\setcounter{page}{1}
\setcounter{figure}{0} 
\setcounter{equation}{0}
\renewcommand{\theequation}{S\arabic{equation}}
\renewcommand{\thefigure}{S\arabic{figure}}
\onecolumngrid

\begin{center}
    {\Large Supplementary Material}
\end{center}

\section*{Matrix element calculation}

The Euclidean finite difference form of the twist-two operator that determines the quark momentum fraction is
\begin{align}
 \label{eq:latticeop}
 {\cal T}_{\mu\nu}^{(\square)}(x)=  
    \frac{1}{4a}\left( \overline{q}(x)\tau_3\gamma_{\{\nu}\large[ U_{\mu\}}(x) q(x+\hat{\mu}) - U^\dagger_{\mu\}}(x-\hat \mu) q(x-\hat \mu) \large]  -\large[\overline{q}(x+\hat\mu) U_{\{\mu}^\dagger(x) - \overline{q}(x-\hat\mu) U_{\{\mu}(x-\hat\mu)\large]\tau_3\gamma_{\nu\}} q(x)\right).
\end{align}
The combination ${\cal T}^{(\square)}= \frac{1}{\sqrt{2}}({\cal T}^{(\square)}_{33}-{\cal T}^{(\square)}_{44})$, which belongs to the $\tau_1^{(3)}$ irreducible representation of  the hypercubic group \cite{Gockeler:1996mu}, is used in this work.
To compute matrix elements of this operator, the compound propagator technique is generalized from that previously used in Refs.~\cite{Savage:2016kon,Chang:2017eiq}.
The operator insertion point is used as a sequential source, and three-point correlation functions are formed by first calculating a (smeared) point-to-all quark propagator extending from the hadronic/nuclear source to the operator insertion point and subsequently calculating additional quark propagators from the operator insertion point to the sink. Since the finite difference form of the operator contains shifts, three different sequential inversions are utilized.
Taking the appropriate linear combinations of displaced sources to implement ${\cal T}^{(\square)}$ results in fixed-order background-field compound propagators that include the operator insertions throughout spacetime.
These compound propagators are used to construct the two-point correlation functions in Eq.~\eqref{eq:G2ptdefBF}.
The background-field two-point correlation functions $G_h(t;\lambda)$ have a spectral representation as a sum of exponentials.
This, in turn, determines the full $t$-dependent form of the ratio of two-point correlation functions in zero and non-zero background fields, $\mathcal{R}_h(t)$ (defined in Eq.~\eqref{eq:R3ptdef}) in terms of the eigenenergies, interpolating operator overlap factors, and ground- and excited-state matrix elements of the lattice operator in Eq.~\eqref{eq:latticeop}.
The ground-state matrix elements of interest for each nuclear system can thus be extracted by fitting LQCD results for  $\mathcal{R}_h(t)$ to the form arising from the spectral representations.

Zero background field two-point correlation function have the spectral representation
\begin{equation}
    G_h^{ss^\prime}(t;\lambda = 0) = \sum_n Z_n^s (Z_n^{s^\prime})^* e^{-E_n t},
    \label{eq:spec}
\end{equation}
where $\{s,s^\prime\} \in \{S,P\}$ specifies the source and sink smearing, $E_n$ is the energy of the $n$-th energy eigenstate, and $Z_n^s$ is an overlap factor defined by $Z_n^s = \sqrt{V}\mbraket{n}{\chi_h^s(0)}{0}$, where $V= a^4\prod_\mu L_\mu$ is the (dimensionful) lattice volume and $L_\mu$ is the extent of the lattice geometry in the $\mu$ direction.
The corresponding background-field two-point functions at ${{\cal O}(\lambda)}$ have the spectral representation
\begin{equation}
\begin{split}
    \left. G_h^{ss^\prime}(t;\lambda)\right|_{{\cal O}(\lambda)} &= a \sum_{\tau = 0}^{t} \sum_{n,m} Z_n^s (Z_m^{s^\prime})^*  e^{-E_n \tau} \mbraket{n}{\tilde T^{(\square)}}{m} \\
    &= \sum_n Z_n^s (Z_n^{s^\prime})^* t\, e^{-E_n t} \mbraket{n}{\tilde T^{(\square)}}{n} \\
    &\hspace{20pt} + \sum_n\sum_{m \neq n} Z_n^s (Z_m^{s^\prime})^* a \left( \frac{e^{-E_n t}}{1- e^{(E_n - E_m)a}} + \frac{e^{-E_m t}}{1- e^{(E_m - E_n)a}} \right) \mbraket{n}{\tilde T^{(\square)}}{m}
    \end{split}
    \label{eq:BGspec}
\end{equation}
where $\tilde T^{(\square)}= \sum_{\bf x} T^{(\square)}({\bf x},0)$ and thermal effects from the finite temporal extent of the lattice have been neglected, see Refs.~\cite{Savage:2016kon,Tiburzi:2017iux,Bouchard:2016heu} for further discussions. The spectral representation for $\mathcal{R}_h^{ss^\prime}(t)$ follows from inserting Eqs.~\eqref{eq:spec} and \eqref{eq:BGspec} into  Eq.~\eqref{eq:R3ptdef} and can be expressed as
\begin{equation}
    \begin{split}
        \mathcal{R}_h^{ss^\prime}(t) &= \sum_n \mbraket{n}{T^{(\square)}}{n} Z_n^s (Z_n^{s^\prime})^* \left(  \frac{ (t+a) e^{-E_n (t+a)}}{\sum_k Z_k^s (Z_k^{s^\prime})^* e^{-E_k (t+a)}} - \frac{t\, e^{-E_n t}}{\sum_k Z_k^s (Z_k^{s^\prime})^* e^{-E_k t}} \right) \\
        &\hspace{20pt} + \sum_{n}  \mathcal{N}_{n}^{ss^\prime} \left(  \frac{ e^{-E_n (t+a)}}{\sum_k Z_k^s (Z_k^{s^\prime})^* e^{-E_k (t+a)}} - \frac{e^{-E_n t}}{\sum_k Z_k^s (Z_k^{s^\prime})^* e^{-E_k t}} \right) 
    \end{split}\label{eq:Rspec}
\end{equation}
where $\mathcal{N}_{n}^{ss^\prime} = a\sum_{m\neq n}  \mbraket{n}{T^{(\square)}}{m} Z_n^s (Z_m^{s^\prime})^*/(1- e^{(E_n - E_m)a}) + \mbraket{m}{T^{(\square)}}{n} Z_m^s (Z_n^{s^\prime})^*/ (1- e^{(E_m - E_n)a})$ involves excited-state transition matrix elements as well as combinations of overlap factors not determined from fits to Eq.~\eqref{eq:spec}. In order to extract the ground-state matrix elements of interest in this work, combined fits to $ G_h^{ss^\prime}(t;0)$ and $\mathcal{R}_h^{ss^\prime}(t)$ are used to fix the free parameters $\{\mbraket{n}{T^{(\square)}}{n},  E_n, Z_n^s (Z_n^{s^\prime})^*, \mathcal{N}_{n}^{ss^\prime} \}$,  for $ss^\prime \in \{ {\rm SS,SP} \}$.
It is noteworthy that if this spectral representation is truncated at $N_{\rm states} = 1$, then the second sum in Eq.~\eqref{eq:Rspec} vanishes and $\mathcal{R}_h^{ss^\prime}(t)$ is independent of $\mathcal{N}_0^{ss^\prime}$.
For $N_{\rm states} > 1$, $\mathcal{R}_h^{ss^\prime}(t)$ is similarly independent from a linear combination of the $\mathcal{N}_n^{ss^\prime}$ that can be used to eliminate one redundant parameter, chosen here to be $\mathcal{N}_0^{ss^\prime}$.
Further, if the sums and finite-difference derivatives in Eq.~\eqref{eq:BGspec} and Eq.~\eqref{eq:Rspec} were replaced by continuum integrals and derivatives, then $\mathcal{R}_h^{ss^\prime}(t)$ would be independent of $\mathcal{N}_n^{ss^\prime}$, suggesting that fits might not be sensitive to $\mathcal{N}_n^{ss^\prime}$ in practice if lattice artifacts are small.
In this work, fits are performed both using Eq.~\eqref{eq:Rspec} and also neglecting these lattice artifacts by setting $\mathcal{N}_n^{ss^\prime} = 0$ in Eq.~\eqref{eq:Rspec}. The Aikiake Information Criterion (AIC) is used to select whether fits with or without these lattice artifacts are preferred for each choice of $N_{\rm states}$ and fit range that is considered.
In all cases, fits without these lattice artifacts are preferred by the AIC and used for subsequent analysis.

\begin{figure*}[!t] 
  \centering
  \includegraphics[width=0.48\textwidth]{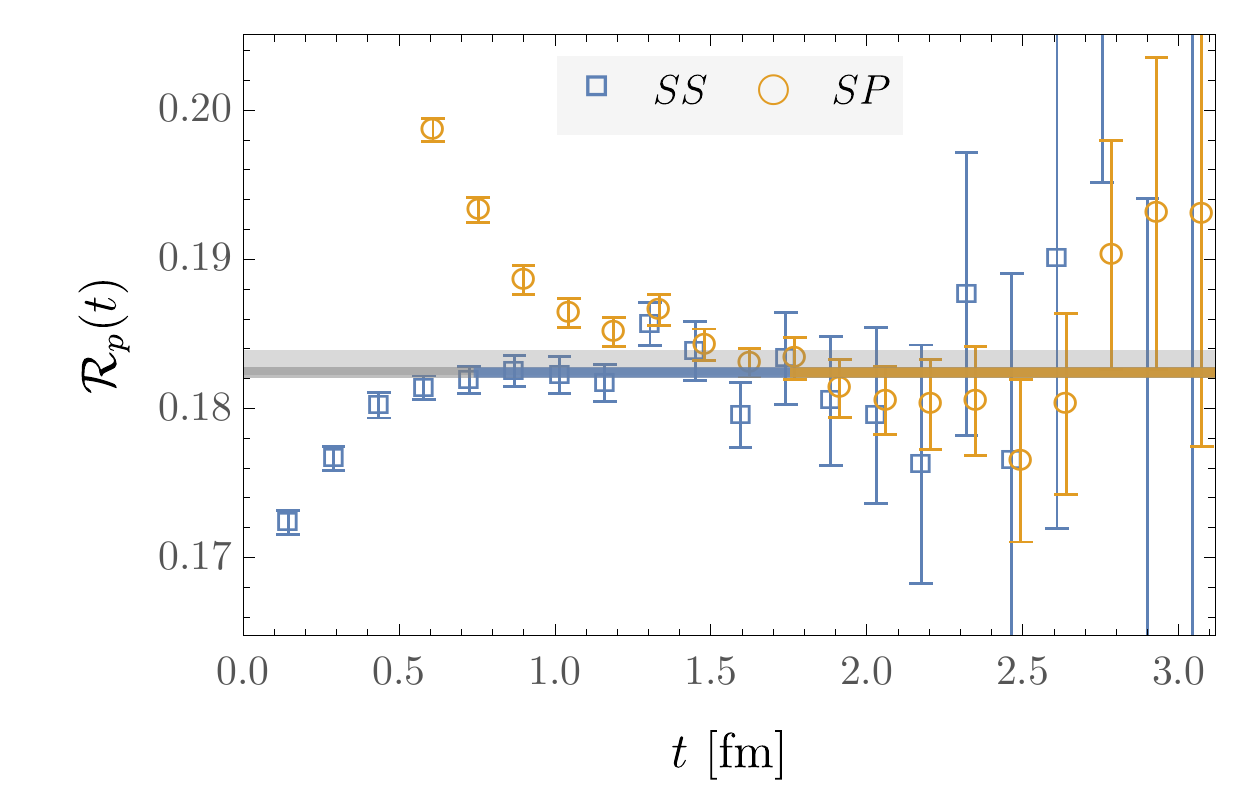}
  \includegraphics[width=0.48\textwidth]{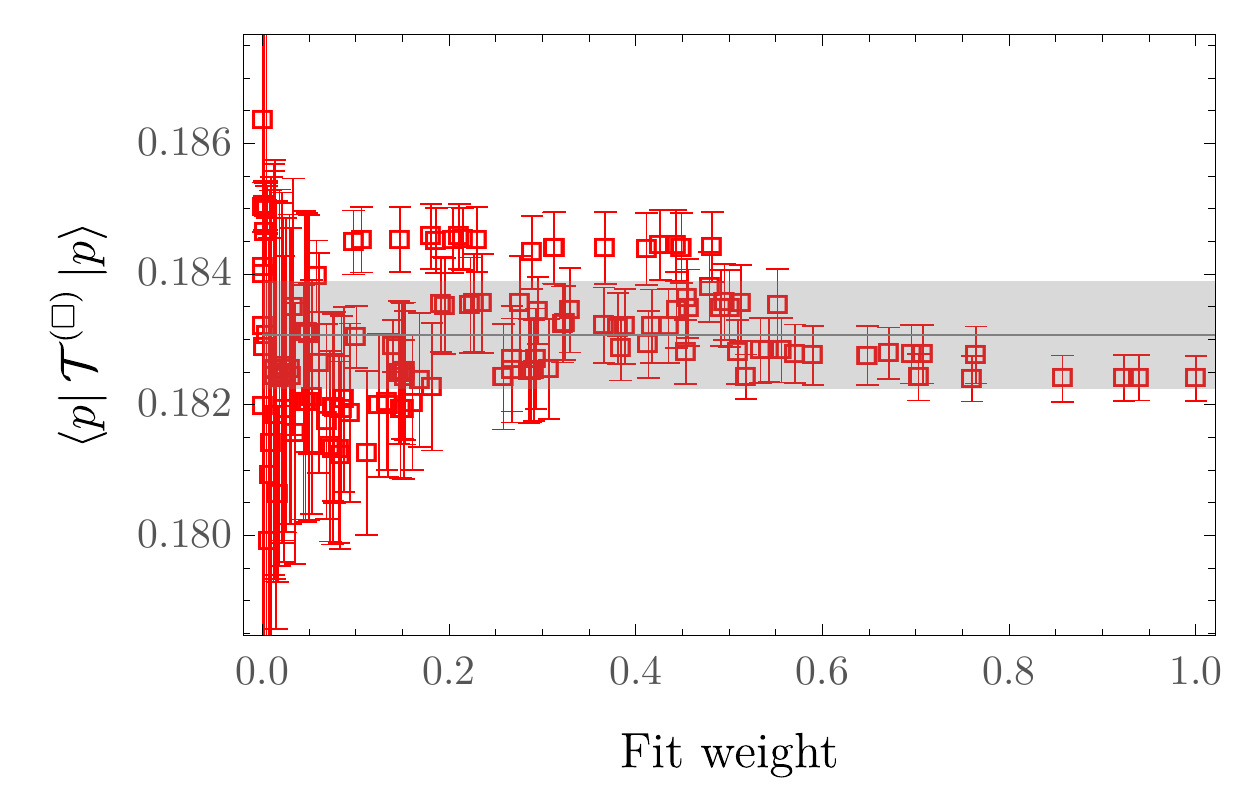}
  \includegraphics[width=0.48\textwidth]{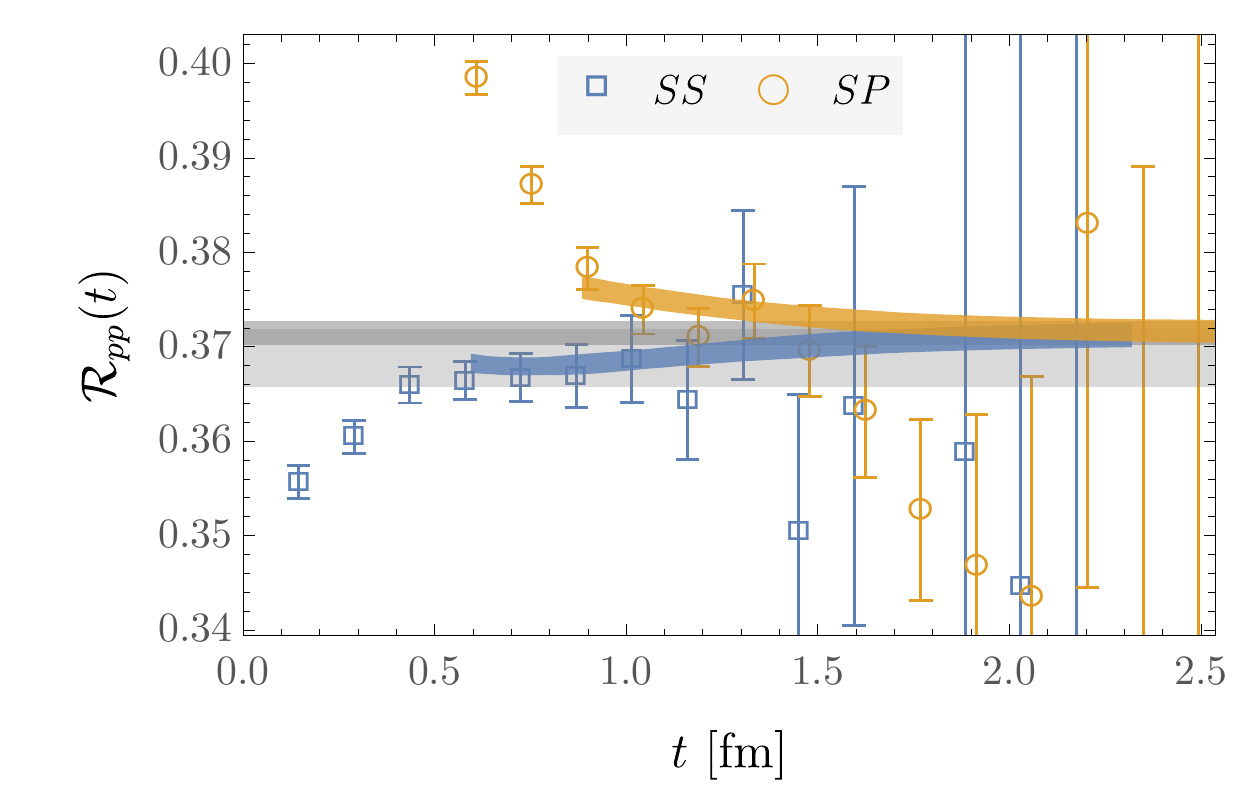}
  \includegraphics[width=0.48\textwidth]{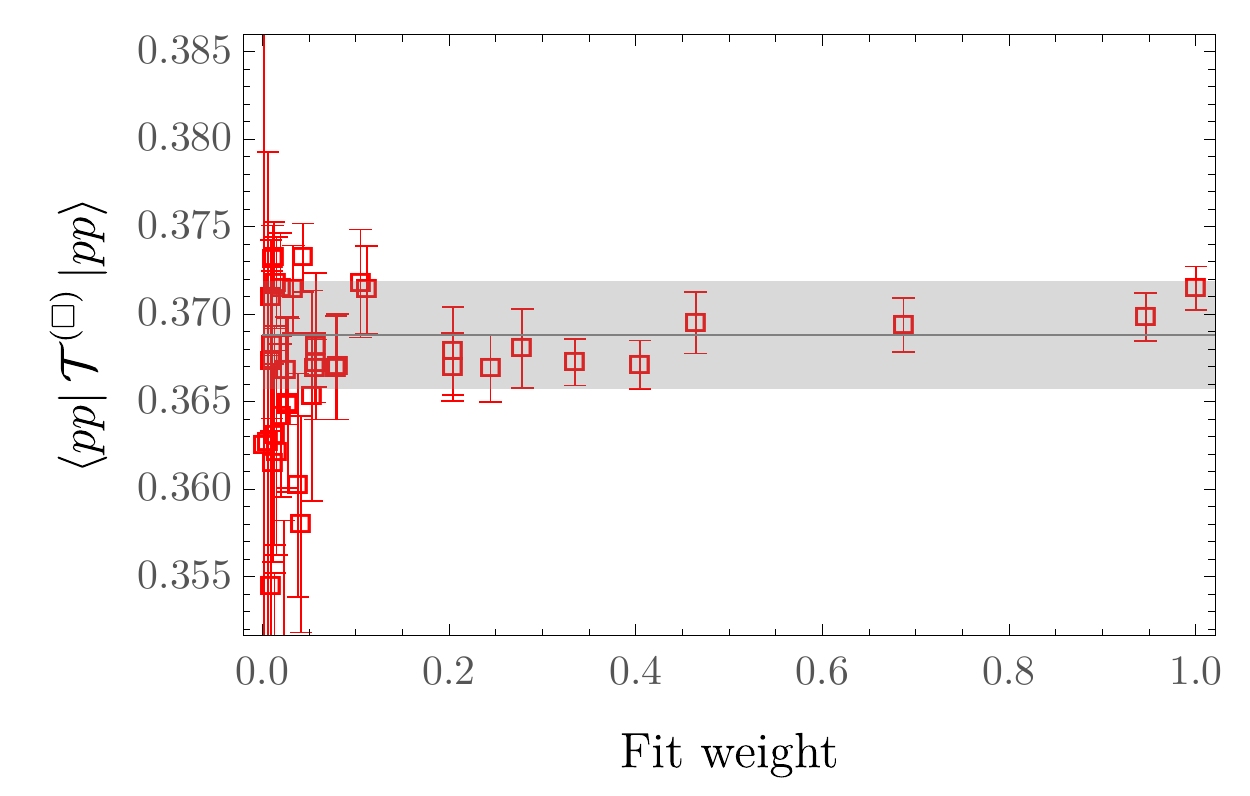}
  \includegraphics[width=0.48\textwidth]{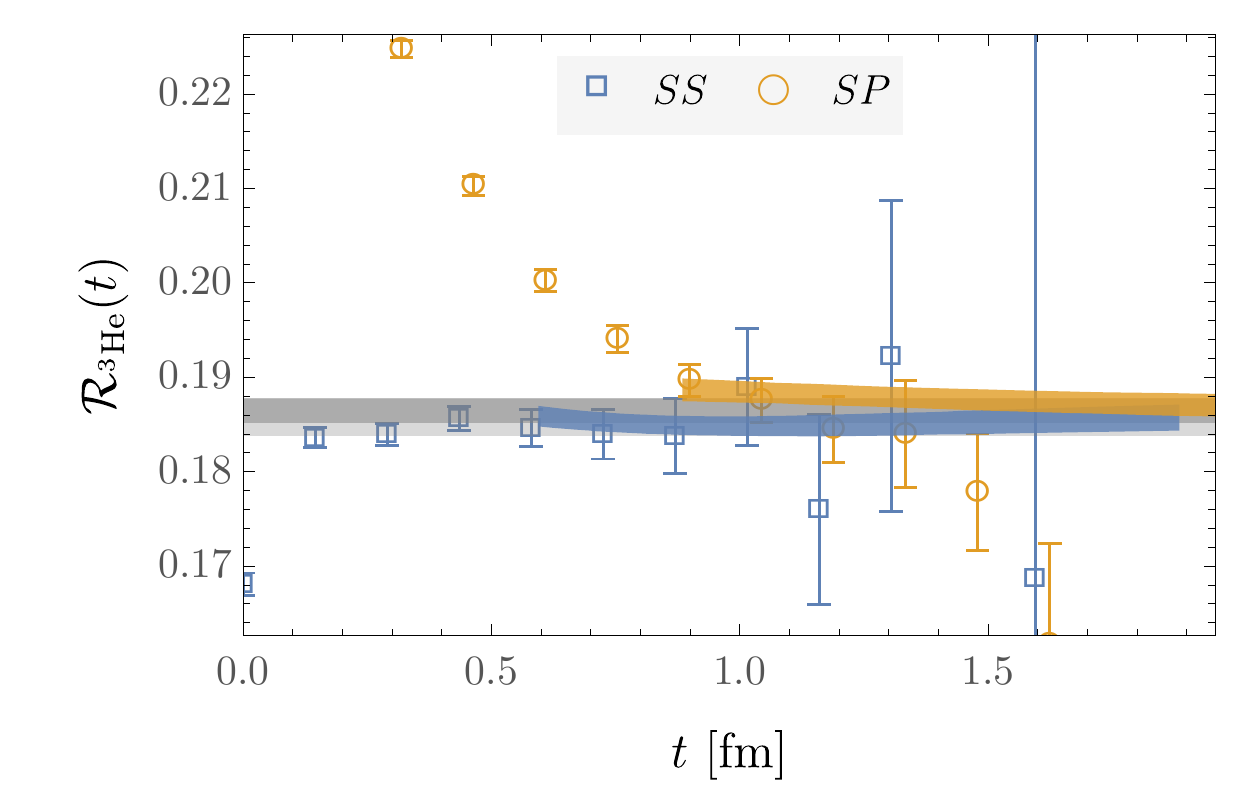}
  \includegraphics[width=0.48\textwidth]{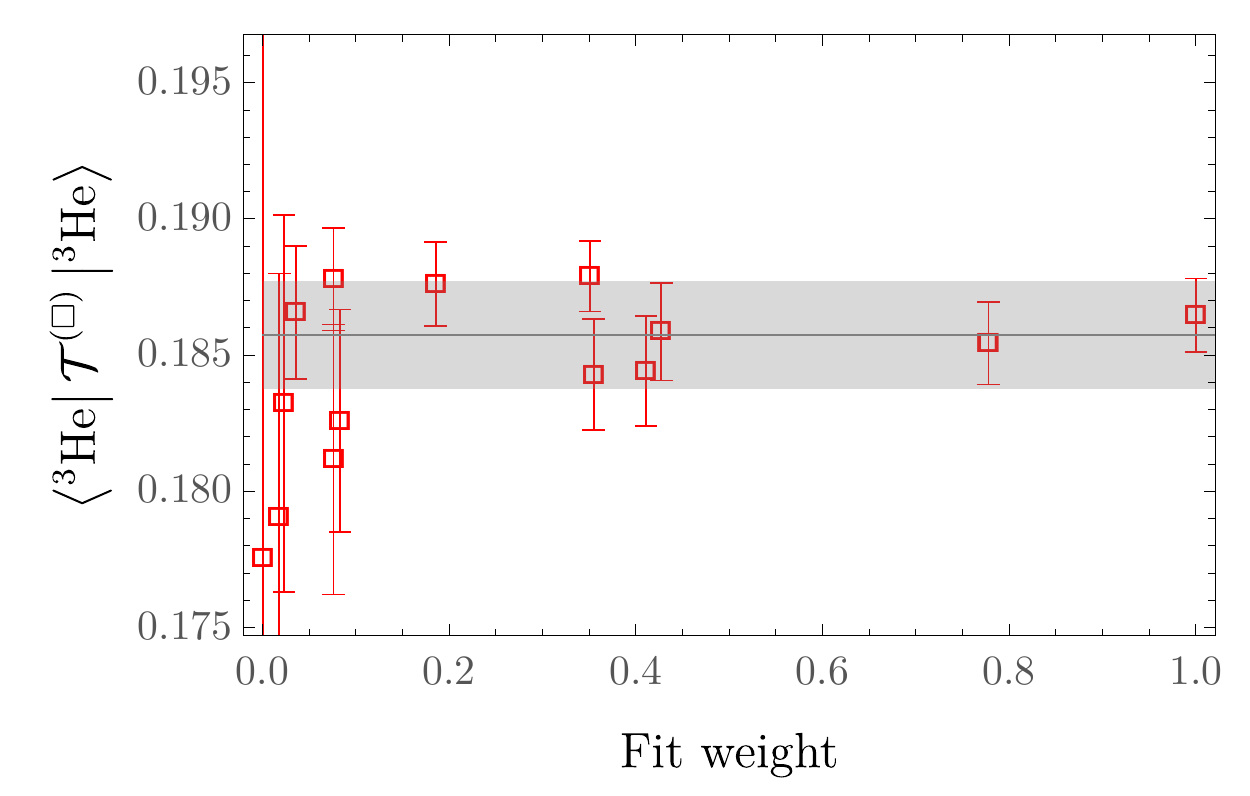}
  \caption{
     Left: Effective matrix elements (Eq.~\eqref{eq:R3ptspec} of the main text), with best fit result (dark grey band) and final fit result (light grey band). Right: The results of all fits included in the weighted average to obtain the systematic and statistical uncertainties of the corresponding matrix element.
    \label{fig:effMEfits}}
\end{figure*}

\begin{figure*}[t!] 
  \centering
  \includegraphics[width=0.48\textwidth]{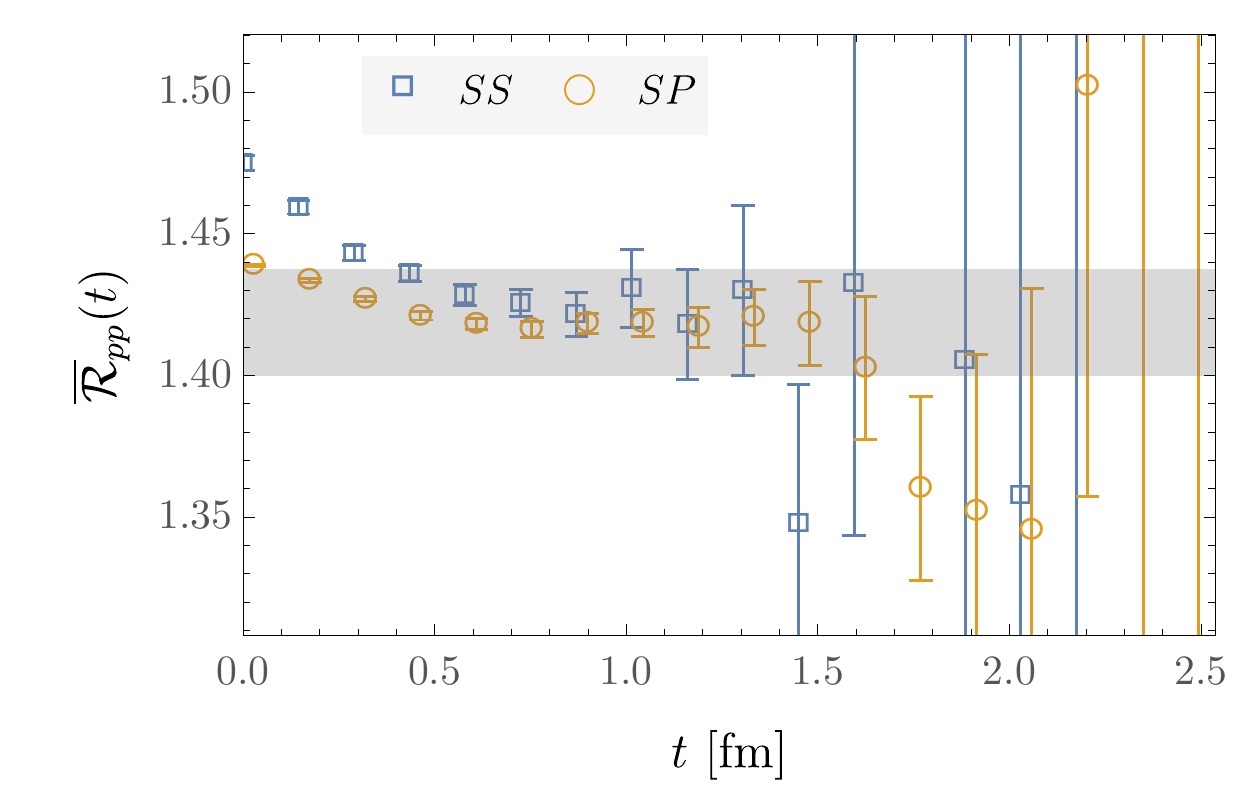}
  \includegraphics[width=0.48\textwidth]{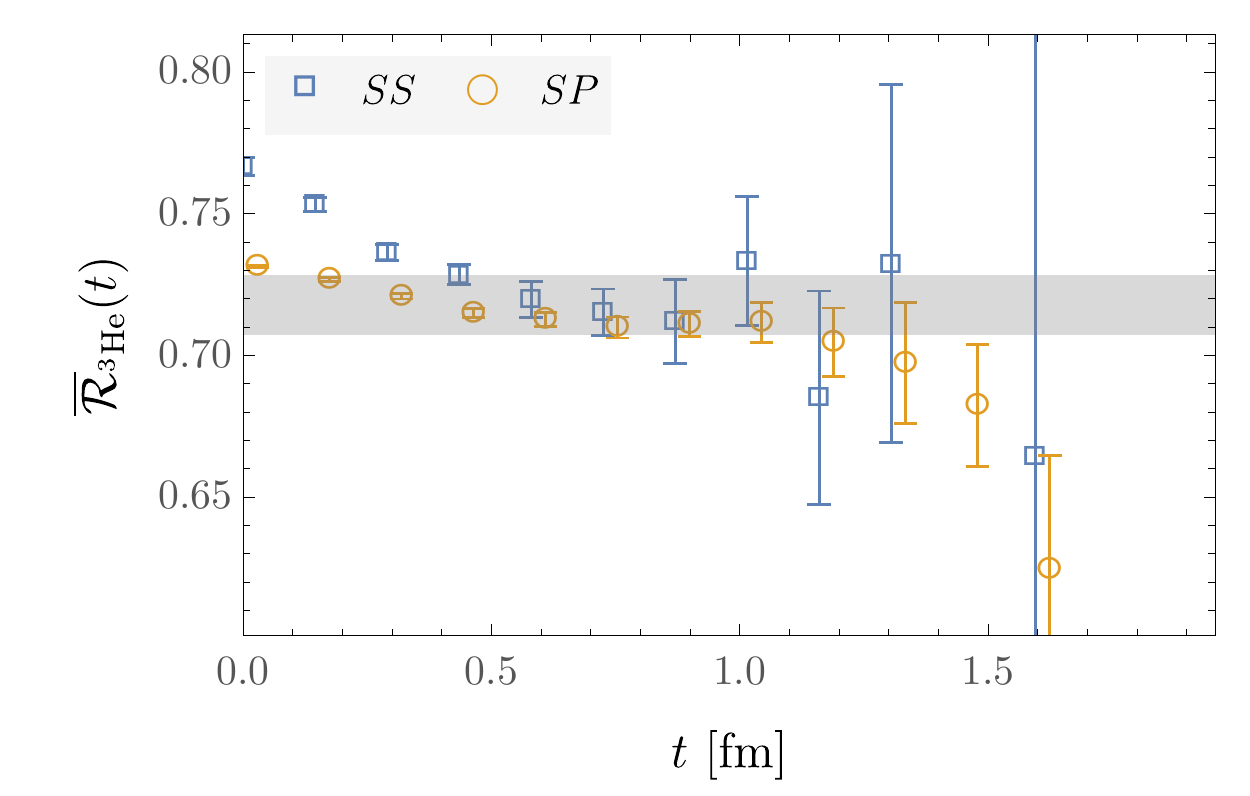}
  \caption{The ratios of effective matrix elements, Eq.~\eqref{eq:Rbar3ptspec}, for $pp$ and $\hethree$ to that in the proton. The gray bands show the reconstructed value of the matrix element ratio from the fits to the individual nuclear states.
    \label{fig:ratfits}}
\end{figure*}

Care must be taken in order to account for systematic uncertainties associated with the fit-range choice and the model selection (i.e., the number of states to retain in truncated approximations to the infinite sums of states in Eqs.~\eqref{eq:spec}-\eqref{eq:Rspec}).
In this work, an automated fitting procedure analogous to the procedure described in Ref.~\cite{Beane:2020ycc} is used to assess these systematic fitting uncertainties.
Results for SS and SP interpolating operators are fit simultaneously and maximum source-sink separations $\{t_{\rm max}^{\rm SS},t_{\rm max}^{\rm SP}\}$ are chosen so that the signal-to-noise ratio of each correlation function is always greater than a tolerance of 0.5.
A minimum separation $t_{\rm min}^{\rm min} = 4$ is required for the spectral representation to be well-defined for three-point correlation functions for the action and operator considered here.
Within these constraints, $N_{\rm max} = 200$ choices of minimum separations for both sources $(t_{\rm min}^{\rm SS},t_{\rm min}^{\rm SP})$ are chosen randomly.
For each fit range, a one-state fit is performed simultaneously for both sources.
A two-state fit is subsequently performed, and if the two-state fit results in an AIC score of $\Delta$AIC$\leq -0.5$, then the two-state fit is accepted.
This procedure is repeated until an $N_{\rm state}$-state fit is rejected, at which point an $(N_{\rm state}-1)$-state fit is used. For the correlation functions analyzed in this work, this procedure results in accepted fits with $N_{\rm state}\in\{1,2,3\}$.
Correlated $\chi^2$-minimization is used for all fits with covariance matrices regularized using optimal shrinkage~\cite{Ledoit:2004,Rinaldi:2019thf}.
The $\chi^2$ depends linearly on the overlap factors, so these parameters are solved for using variable-projection  methods~\cite{Golub:2003,Rust:2013}.
The excited-state energy gaps are then determined using nonlinear optimization (Nelder-Mead and gradient-based Newton solvers from the \texttt{Julia} package \texttt{Optim}  are used and verified to reproduce the same minimum energy gaps within an absolute tolerance of $10^{-5}$).
The fit is then repeated for $N_{\rm boot}=200$ bootstrap samples, and the marginalized parameter uncertainty on the ground-state matrix element is determined using rank-based bootstrap confidence interval estimation~\cite{Davison:1997} to add robustness to outlier bootstrap samples.
Various checks on the fit result are then performed: an uncorrelated fit must reproduce the result within a tolerance of 5-sigma, the bootstrap median must reproduce the mean within a tolerance of 2-sigma, and the goodness-of-fit must be below a tolerance of $\chi^2/N_{\text{dof}} < 2$.
All fit results passing these criteria are included in an ensemble of acceptable fits.
A weighted average of these acceptable fits is used to determine the final central value and total systematic uncertainty, where the (in principle arbitrary - see Ref.~\cite{Jay:2020jkz} for a discussion of this in a Bayesian framework) weights are taken to be the ratio of the $p$-value to the variance of each fit as in Ref.~\cite{Rinaldi:2019thf}. Final uncertainties are obtained by adding in quadrature the statistical uncertainty of the highest weight fit with the systematic uncertainty obtained from the weighted average.
The fitting procedure is fully specified by the parameters and tolerance values above, as well as by a random seed for bootstrap resampling that is fixed to allow correlated ratios of matrix elements for different hadrons to be formed.

Fig.~\ref{fig:effMEfits} shows summaries of the  fits of $\mathcal{R}_h^{\rm 3pt}$ for each state studied in this work using the approach described above.  It is also convenient to define the further ratio for a nuclear state $h$:
\begin{equation}
     \overline{\mathcal{R}}_h(t) =  \mathcal{R}_h(t) / \mathcal{R}_p(t) 
     \label{eq:Rbar3ptspec}
 \end{equation}
which determines the ratio of the nuclear momentum fractions to that of the proton. Fig.~\ref{fig:ratfits} shows results for $\overline{\mathcal{R}}_h(t)$ including results obtained by fitting $\mathcal{R}_h(t)$ and $\mathcal{R}_p(t)$ independently as described above, using correlated bootstrap resampling to determine ratios of the ground-state matrix elements from all successful pairs of fit ranges, and taking a weighted average of the results.

\section*{Nonperturbative renormalization}

An ensemble of $N_{\rm meas}=101$ field configurations is used to compute the nonperturbative renormalization of the operator studied in this work. The parameters of the ensemble match those used for the nuclear matrix element calculations in the main text, except they have a smaller volume of $L^3\times T=24^3\times24$.

The renormalization coefficient of the local operator $\mathcal{T}_{\mu\nu}$ in Eq.~\eqref{eq:Tqlattdef} is computed in the $\rm{RI'MOM}$ scheme~\cite{Martinelli:1994ty} by 
equating the  Landau-gauge dressed vertex function
\begin{equation}
	\Gamma_{\mu\nu}(p)\equiv\frac{1}{N_C} {\rm Tr}_C\left[ S^{-1}(p) G_{\mu\nu}(p) S^{-1}(p) \right]
\end{equation} 
with its tree level value at a fixed momentum $p$. Here, $S$ is the quark propagator and $G_{\mu\nu}$ is the quark three-point correlation function for the zero-momentum projected operator 
$\tilde{\mathcal{T}}^{(\square)}_{\mu\nu}$, and the trace is over color degrees of freedom. Defining $\tilde{\mathcal{T}}^{(\square)}_{\mu\nu}$ =
$\sum_z\mathcal T^{(\square)}_{\mu\nu}(z) = \sum_{z, z'}\overline q(z)\, J^{(\square)}_{\mu\nu}(z, z')\,q(z')$, 
this zero-momentum-projected three-point correlation function is
\begin{equation}
	G_{\mu\nu}(p) = \frac{1}{V}\sum_{x, y, z, z'}e^{ip(x - y)} S(x, z) J^{(\square)}_{\mu\nu} (z, z') S(z', y),
	\label{eq:greens_function}
\end{equation}
which is computed with the sequential source technique applied through the operator~\cite{Martinelli:1988rr}.
At tree-level, the vertex function  is proportional to two tensor structures~\cite{Gracey:2003mr}:
\begin{align}
	i\Lambda_{\mu\nu}^{1}(p) \equiv \frac{1}{2}(\tilde p_\mu \gamma_\nu + \tilde p_\nu \gamma_\mu) - \frac{1}{4} \slashed{\tilde p} \delta_{\mu\nu}, \hspace{1cm}
	i\Lambda_{\mu\nu}^2(p) \equiv \frac{\tilde p_\mu \tilde p_\nu}{\tilde p^2} \slashed{\tilde p} - \frac{1}{4} \slashed{\tilde p} \delta_{\mu\nu},
\end{align}
where $\tilde p_\mu\equiv \frac{2}{a}\sin\left(\frac{a}{2}p_\mu\right)$  is the lattice momentum corresponding to $p_\mu$. In the continuum, only $\Lambda^{1}_{\mu\nu}$ appears, however $\Lambda_{\mu\nu}^{2}$ 
enters as an $\mathcal O(a)$ correction~\cite{Gracey:2003mr}.
Expressing $\Gamma_{\mu\nu}$ in the space spanned by $\{\Lambda^{(1)}, \Lambda^{(2)}\}$ amounts to imposing the 
renormalization condition:
\begin{equation}
	\left.\Gamma_{\mu\nu}(p)\right|_{p^2 = \mu_0^2} = \left. \mathcal{Z}_q(p)\left[\mathcal{Z}_1^{-1}(p) \Lambda^1_{\mu\nu}(p) + \mathcal{Z}_2^{-1}(p) \Lambda^2_{\mu\nu}(p)\right]\right|_{ p^2 = \mu_0^2}\,,
	\label{eq:mix_rc}
\end{equation}
where $\mathcal{Z}_a^{-1}(p)$ for $a\in\{1,2\}$ are operator renormalization coefficients and the quark 
field renormalization $\mathcal Z_q$ is defined as:
\begin{equation}
	\mathcal Z_q({p})|_{p^2 = \mu_0^2} = \frac{i}{12\tilde p^2}\textnormal{Tr}\left[S^{-1}( p)\slashed{\tilde p}\right]\bigg|_{p^2 = \mu_0^2}.~
	\label{eq:quark_renorm}
\end{equation}
The renormalization coefficient of primary interest is $\mathcal Z^{\rm{RI'MOM}}( \mu_0,a)\equiv \mathcal Z_1(p)\big|_{p^2=\mu_0^2}$.

Eq.~(\ref{eq:mix_rc}) is solved by constructing a linear functional on the space of Dirac and Lorentz 
matrices. A bilinear form $\langle\cdot, \cdot\rangle$ is defined whose action is:
\begin{equation}
	\langle \lambda^1, \lambda^2\rangle \equiv \sum_{i\in\tau_1^{(3)}} \textnormal{Tr}_\textnormal{D}\left\{\lambda^1_i 
	\lambda^2_i\right\},
\end{equation}
where $\lambda^1, \lambda^2$ are objects with two Dirac and two Lorentz indices and the sum runs over elements of a basis for the irreducible representation of the hypercubic group that contains $\mathcal{T}$, as defined in Eq.~\eqref{eq:Tqlattdef}. The functionals $\langle\Lambda^1, \cdot\rangle$ and 
$\langle\Lambda^2, \cdot\rangle$ are applied to Eq.~(\ref{eq:mix_rc}) to yield a system of equations:
\begin{equation}
{\cal Z}_q(p)	\sum_{b\in\{1,2\}} \langle \Lambda^a(p), \Lambda^b (p)\rangle{\cal Z}_b^{-1}(p) = \langle \Lambda^a(p), \Gamma(p) \rangle,
	\label{eq:rc_soln}
\end{equation}
which can be solved for ${\cal Z}_{1,2}({p})$.

\begin{figure}[t]
	\centering
	\includegraphics[width=0.6\textwidth]{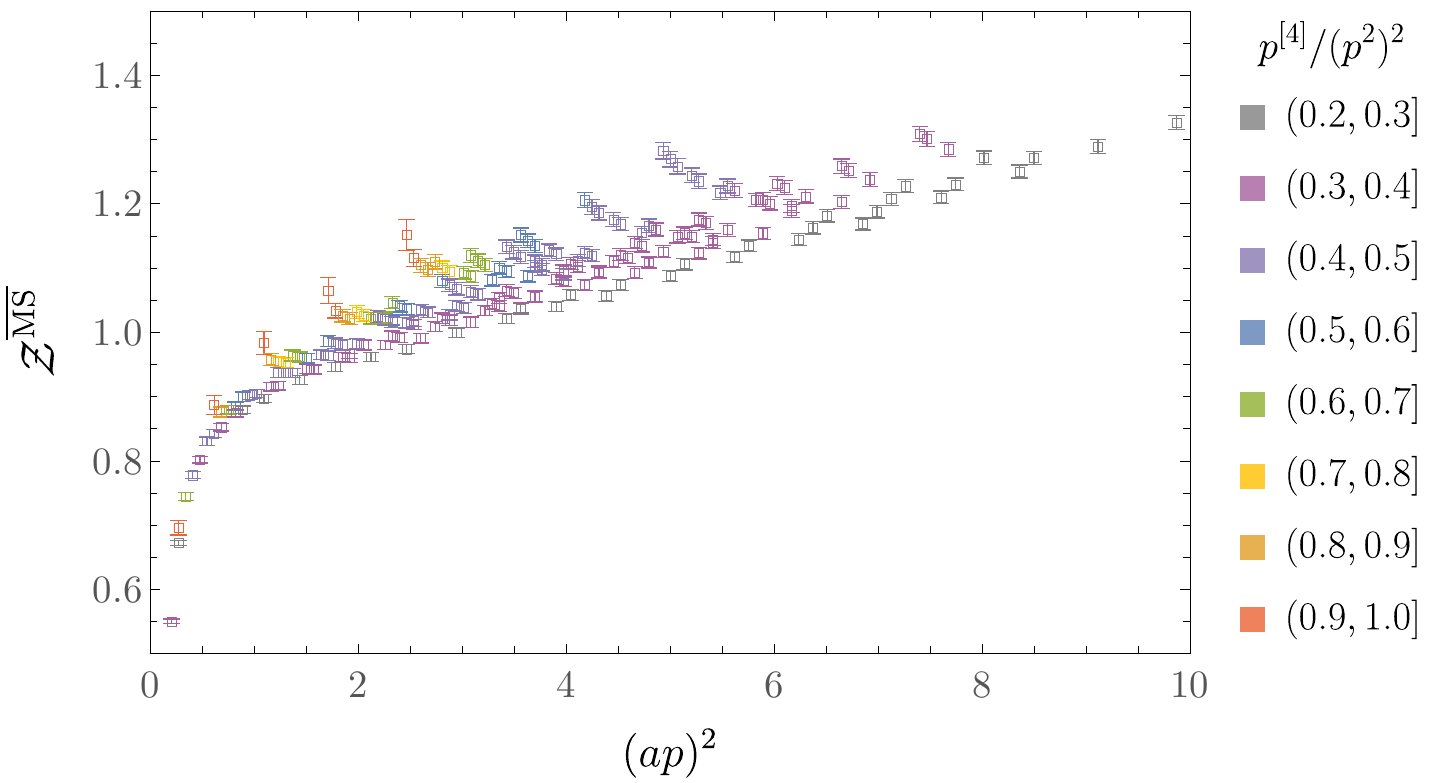}
	\caption{Renormalization coefficients in the $\overline{\textnormal{MS}}$ scheme, color-coded by the amount of hypercubic breaking at 
	each point. $\mathcal{Z}^{\overline{\textnormal{MS}}}$ is computed 
	on each ensemble for each mode 
	$p_\mu$ with $0.5 \leq (ap)^2 \leq 10$,
	and averaged over hypercubic orbits.}
	\label{fig:msbar}
\end{figure}

An $\overline{\textnormal{MS}}$ renormalization factor can be constructed from the $\rm{RI'MOM}$ factor computed as described above as 
\begin{align}\label{eq:ZMSbar}
	\mathcal{Z}^{\overline{\textnormal{MS}}}(\mu) &= 
	\mathcal{R}^{\overline{\rm MS}/\rm{RI'MOM}}(\mu,\mu_0)	\mathcal{Z}^{\rm{RI'MOM}}(\mu_0,a),
\end{align}
where the matching factor $\mathcal{R}^{\overline{\rm MS}/\rm{RI'MOM}}(\mu,\mu_0)$ has been computed to 3-loop order in lattice perturbation theory~\cite{Gracey:2003yr,Gracey:2003mr}
(using the two-loop value produces a statistically indistinguishable result).
While $\mathcal{Z}^{\overline{\textnormal{MS}}}(\mu)$ defined in Eq.~\eqref{eq:ZMSbar} is in principle independent of the matching scale $p^2=\mu_0^2$, in practice there are discretization artifacts which arise as contamination in the form of dependence on the hypercubic invariants $p^{[2n]}\equiv \sum_\mu p_\mu^{2n}$, resulting in the classic jellyfish-bone structure shown in Fig.~\ref{fig:msbar}. $\tilde p_\mu$ can be expanded in a Taylor series in $p^{[2n]}$, so either $p^{[2n]}$ or $\tilde p^{[2n]}$ may be used to perform the fit; $p^{[2n]}$ is used for this analysis, but consistent results are obtained by fitting to $\tilde p^{[2n]}$.

To obtain the renormalization factor, the hypercubic artifacts which depend on $p^{[2n]}$ for $n > 1$ are fit using the one-window-fit approach, detailed in Ref.~\cite{Blossier:2010vt}. Running terms and remaining artifacts which depend only on $p^2$ are then fit separately. 
To allow a quantification of systematic uncertainties, the fits are performed over a range of windows, with a range of functional forms.
In particular, artifacts of the form
\begin{align}
	\left\{H_1, H_2, H_3\right\} &= \left\{c_1\frac{a^2p^{[4]}}{p^2} + c_2 \frac{a^2p^{[6]}}{(p^2)^2}, c_3\frac{a^2p^{[4]}}{p^2} \log(a^2p^2),	c_4 a^4 p^{[4]} + c_5  \frac{a^4p^{[6]}}{p^2} + c_6 a^4\left(\frac{p^{[4]}}{p^2}\right)^2 \right\} \\
	\left\{R_1, R_2, R_3, R_4\right\}&= \left\{d_1 a^2p^2, \frac{d_2}{a^2p^2}, d_3 \,a^2p^2 \log(a^2p^2), d_4 \left(a^2p^2\right)^2\right\},
\end{align}
are considered, where the $H_i$ denote hypercubic terms and the $R_i$ are contributions from running. Terms are grouped by their order in $a^2$ and by whether or not they contain logarithmic corrections.
The full functional form is truncated at order $a^4$. Adding further logarithmic terms into the functional form does not increase the fit quality. 
A fit form $F$, constructed from these components, is chosen for a specific data window to maximize the goodness of fit while preventing overfitting using the following procedure.
Given a window of momenta, the first fit form is initialized to be $F^{(1)} = 
0$. Given a fit form $F^{(n)}$, the subsequent form $F^{(n + 1)}$ is determined by considering all possible forms
\begin{equation}
	F_j^{(n + 1)} = F^{(n)} + X_j,
\end{equation} 
which can be built from $F^{(n)}$ using only the terms $X_j$ which are not currently present in $F^{(n)}$, and $X\in\{H,R\}$ is as appropriate for the fit to the hypercubic or running artifacts. 
A fit form $F_j^{(n + 1)}$ is accepted if and only if $A(F_j^{(n + 1)}) < A(F_j^n)$, where $A(F)
\equiv 2 N_\textnormal{param}(F) + \chi^2(F)$ is the value of the AIC of the fit $F$. 
If no forms are accepted or there are no fit forms left, iteration stops and $F\equiv F^{(n)}$.
The subsequent fit form $F^{(n + 1)}$ is chosen out of the accepted fit forms $\{F_j^{(n + 1)}\}_j$ by maximizing the $p$-value of the fit. 

This procedure is applied to a range of fitting windows. Windows are chosen with $(ap)^2\in [(ap)^2_{\textnormal{min}}, 
(ap)^2_{\textnormal{max}}]$ and $p^{[4]} / (p^2)^2\leq h_0$ by independently adjusting $(ap)^2_{\textnormal{min}}$ and 
$(ap)^2_{\textnormal{max}}$ between 0.5 and 10 in increments of 0.5, always keeping a window size $(ap)^2_{\textnormal{max}} - 
(ap)^2_{\textnormal{min}}\geq 5$, and taking $h_0\in \{0.5, 0.6, ..., 1.0\}$. For each fit window $f$, the optimal fit form as defined above is accepted 
if and only if its $p$ value $p_f\geq 0.01$, and each accepted fit is given a weight:
\begin{equation}
	w_f\equiv p_f (\delta\mathcal{Z}_f^{\overline{\text{MS}}})^{-2},
\end{equation}
where $\delta\mathcal{Z}^{\overline{\text{MS}}}_f$ denotes the statistical uncertainty on $\mathcal{Z}_f^{\overline{\text{MS}}}$, which is the fit result for $\mathcal{Z}^{\overline{\text{MS}}}$ that is obtained in fit $f$. The weights are normalized so that the maximum weight is 1.
Given the large number of accepted fits, the $N_w\equiv 100$ highest-weight fits are combined in a weighted average, with the result:
\begin{equation}
	\mathcal{Z}^{\overline{\textnormal{MS}}}(\mu=2\ \text{GeV}) = 0.885(42).
\end{equation}
The final result and uncertainties are consistent under variation of $N_w$, with larger systematic uncertainties when additional fits with lower weights are included (e.g., with $N_w = 200$, the result is 0.895(55)).
\begin{figure}[t]
\centering
	\includegraphics[width=.6\textwidth]{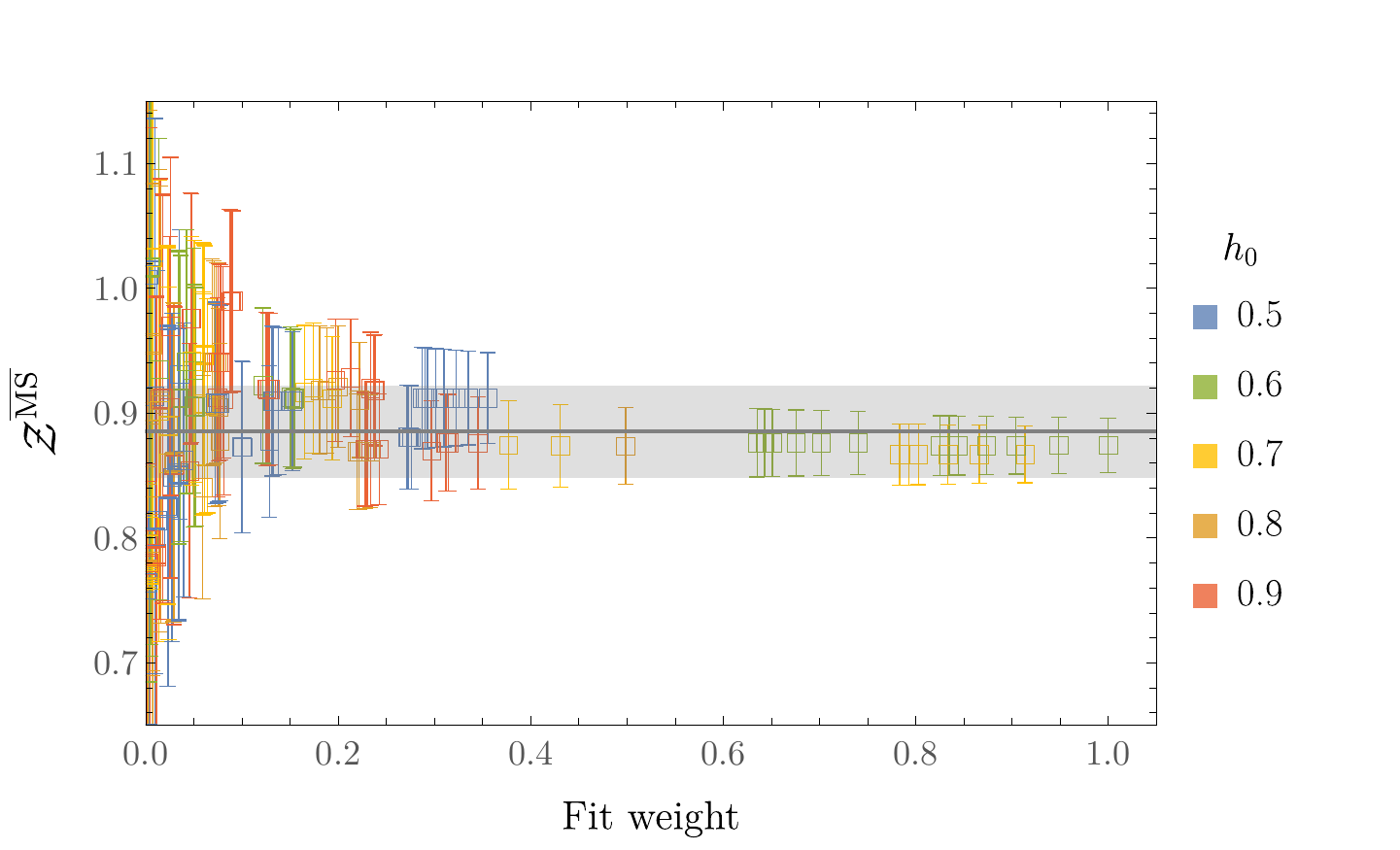}
	\caption{$N_w=100$ highest-weight accepted fits, ${\cal Z}_f$, plotted as a function of their fit weight, $w_f$. Individual fits are color-coded by the maximum amount of hypercubic 
	breaking in the window, and the gray band shows the final weighted average.}
	\label{fig:fits}
\end{figure}
As a consistency check, the computation was also performed on an  ensemble with identical action and parameters with lattice dimensions $16^3\times 48$. A result of $\mathcal{Z}^{\overline{\textnormal{MS}}}(\mu=2\ \text{GeV})=0.910(57)$ is obtained, indicating finite volume effects are negligible within uncertainties.

\section{Connection to phenomenology: Momentum fraction extrapolation}

In order to connect to phenomenology, the  lattice result for $\langle x\rangle_{u-d}^{(\hethree)}$ at quark masses corresponding to $m_\pi=806$ MeV is extrapolated to the physical masses using the assumption of weak mass dependence of the short-distance two-body counterterms in nuclear EFT. 
The isovector twist-two operators in Eq.~\eqref{eq:opdef} match on to hadronic operators in nuclear EFT as \cite{Chen:2004zx}
\begin{equation}
    {\cal O}_{\mu_1\ldots\mu_n}\rightarrow
    \langle x^n \rangle^{(p)}_{u-d} v_{\mu_1} \ldots v_{\mu_n} N^\dagger \tau_3 N (1+\alpha_{3,n} N^\dagger N) +\ldots
\end{equation}
where $N$ is a nucleon field, $v$ is the nuclear velocity, and the ellipsis denotes higher order nucleonic and pionic operators.
Defining the nuclear factor ${\cal G}_3(N,Z)=\langle N,Z|N^\dagger \tau_3 N N^\dagger  N|N,Z\rangle$, the two-nucleon counterterm $\alpha_{3,2}$ relates the nuclear and proton momentum fractions as~\cite{Chen:2004zx}
\begin{equation}\label{eq:counterterm1}
    \alpha_{3,2}{\cal G}_3(N,Z) \equiv \langle x \rangle_{u-d}^{(N,Z)} - \frac{Z-N}{A}\langle x \rangle^{(p)}_{u-d}.
\end{equation}
This LEC can be determined from the LQCD results for $\hethree$ most precisely by re-expressing it in terms of the quantities in Table~\ref{tab:bare} as
\begin{equation}\label{eq:counterterm}
    \alpha_{3,2}{\cal G}_3(\hethree) =\frac{1}{3}\left(3\frac{ \langle x \rangle_{u-d}^{(\hethree)}}{\langle x \rangle_{u-d}^{(p)}} -1 \right)\langle x \rangle^{(p)}_{u-d}.
\end{equation}
Since renormalization effects cancel in the ratio, it is more precisely determined than the individual momentum fractions themselves and, with a naive error propagation, computing $\alpha_{3,2}{\cal G}_3(\hethree)$ via Eq.~\eqref{eq:counterterm} rather than Eq.~\eqref{eq:counterterm1}  achieves smaller uncertainties. 

The matching of the LEC, and extrapolation to physical quark masses proceeds in the following steps:
\begin{enumerate}
    \item Determine the counterterm $\alpha_{3,2}{\cal G}_3(\hethree)$ at $m_\pi=806$~MeV via Eq.~\eqref{eq:counterterm}, using the LQCD calculations of the ratio of $\langle x\rangle_{u-d}^{(\hethree)}/\langle x\rangle_{u-d}^{(p)}$, and $\langle x\rangle_{u-d}^{(p)}$ itself.
 
    \item While the momentum fractions themselves have nonanalytic quark mass dependence, the counterterms $\alpha_{3,2}$ and nuclear factors ${\cal G}_3(N,Z)$ are expected to have only mild quark mass dependence and so their their combination is extrapolated to the physical quark masses by assuming the same central value and increasing the uncertainty by 50\%. 

    \item The extrapolated value of $\alpha_{3,2}{\cal G}_3(\hethree)$ is combined with the value of $\langle x\rangle_{u-d}^{(p)}$ from phenomenology to produce a physical-point value of $\langle x\rangle_{u-d}^{(\hethree)}/\langle x\rangle_{u-d}^{(p)}$.
\end{enumerate}

\end{document}